\documentclass[final,5p,times,twocolumn,monochrome]{elsarticle} 

\usepackage{graphicx}
\usepackage{amssymb}
\usepackage{amsmath}

\usepackage[modulo]{lineno}

\usepackage{color}
    
\journal{Journal Name}

\begin{document}

\begin{frontmatter}

\title{Acceleration of cometary dust near the nucleus:  
Application to 67P/Churyumov-Gerasimenko.}

\author[1]{Yuri Skorov}
\author[2,3]{Volodymyr Reshetnyk}
\author[1]{Pedro Lacerda}
\author[1]{Paul Hartogh}
\author[4]{J\"urgen Blum}

\address[1]{Max-Planck-Institut f\"ur Sonnensystemforschung, Justus-von-Liebig-Weg 3, D-37077 G\"ottingen, Germany}
\address[2]{Taras Shevchenko National University of Kyiv, Glushkova ave. 2, Kyiv, Ukraine}
\address[3]{ Main Astronomical Observatory of National Academy of Science of Ukraine, Akademika Zabolotnoho Str. 27, 03680 Kyiv, Ukraine}
\address[4]{Technische Universit\"at Braunschweig, Institut für Geophysik und extraterrestrische Physik, Mendelssohnstra{\ss}e 3, D-38106 Braunschweig, Germany}

\begin{abstract}

We present a model of cometary dust capable of simulating the dynamics within the first few tens of km of the comet surface. Recent measurements by the GIADA and COSIMA instruments on Rosetta show that the nucleus emits fluffy dust particles with porosities above 50\% and sizes up to at least mm \citep{Schulz,Rotundi,Fulle2015}. Retrieval of the physical properties of these particles requires a model of the effective forces governing their dynamics.  Here, we present a model capable of simulating realistic, large and porous particles using hierarchical aggregates, which shows previous extrapolations to be inadequate. The main strengths of our approach are that we can simulate very large (mm-scale) non-spherical agglomerates and can accurately determine their 1) effective cross-section and ratio of cross-section to mass, 2) gas drag coefficient, and 3) light scattering properties. {\color{red} In practical terms, we find that a more detailed treatment of the dust structure results in 3-5 times higher velocities for large dust particles in the inner coma than previously estimated using spherical particles of the same mass.} We apply our model to the dynamics of dust in the vicinity of the nucleus of comet 67P and successfully reproduce the dust speeds {\color{red} reported early on when the comet was roughly 3.5 AU from the Sun}. At this stage, we employ a simple spherical comet nucleus, we model activity as constant velocity gas expansion from a uniformly active surface, and use Mie scattering. We discuss pathways to improve on these simplifications in the future.

\end{abstract}


\end{frontmatter}


\section{Introduction}
\label{Introduction}

Cometary activity, i.e.\ the release of gas and dust from cometary nuclei as they approach the inner regions of the solar system is what makes comets unique objects in the solar system. Ice sublimation due to nucleus heating by solar radiation triggers the accompanying release of a non-volatile component (cometary dust) and leads to the formation of the gas-dust coma. Despite their key importance in the study of comets, these processes remain poorly understood.  
 
Most information on cometary activity comes from numerous ground-based photometric, spectroscopic and polarimetric observations of the coma and tails of comets. Supported by sophisticated theoretical models, these observations have led to a basic understanding of the microphysical properties and composition of cometary dust, as well as the large-scale, spatial dynamic structure of the dust coma and tails. We point the reader to the book Comets II\footnote{Comets II (Space Science Series), 2004 by Michel C. Festou (Editor), H. Uwe Keller (Editor), Harold A. Weaver (Editor)} for a comprehensive review of the state of art of the subject.  
 
Most attempts to explain the dynamics of the dust coma seen in ground-based observations consider only dust motion at large cometocentric distances ($\sim$$10^6$ m). There, the nucleus gravity is negligible and the dust dynamics is dominated by solar gravity and light pressure. In contrast, models investigating the complex inner coma, which consider only the forces due to the cometary nucleus (gravity and gas drag) are poorly constrained by observations. Moreover, in all of the above the structure of dust particles is often oversimplified, e.g., the dust is treated as solid spheres with a limited size range. 
 
The Rosetta mission offers a unique opportunity to investigate the properties of cometary dust and how they influence the dynamics of the coma at a broad range of scales. Here we focus on results from mainly three instruments on-board Rosetta: the Cometary Secondary Ion Mass Analyser (COSIMA) instrument on-board Rosetta, which collects dust grains from the near-nucleus environment onto target plates for subsequent imaging and compositional analysis, the Grain Impact Analyser and Dust Accumulator (GIADA), which provides the speed, momentum and approximate cross-section of individual grains down to approximately 10 $\mu$m in size, and the Optical, Spectroscopic, and Infrared Remote Imaging System (OSIRIS), which is the scientific camera system on {\color{blue} the orbiter}. Three interesting properties of the grains collected by COSIMA and GIADA are \citep{Rotundi,Schulz,Fulle2015}: 
\begin{enumerate}
\item the grains appear weakly bound, fluffy agglomerates with porosities about 50\%;
\item large grains, up to several hundred microns in size have been collected;
\item the grains hit the detectors at speeds 1 to 10 m/s, and their speed is not strongly correlated with size. 
\end{enumerate}

{\color{red}The OSIRIS camera images taken at heliocentric distances above 3.6 AU reveal tracks of particles up to 1.7 cm in diameter \citep{Rotundi}.
\citet{Fulle2015} found that the dust particles impacting on GIADA detectors can be separated into two families: smaller `compact' particles, and somewhat larger `fluffy aggregates'. It should be noted that even the `compact' particles  do not appear solid and have considerable porosity.
In \citep{Fulle2015b}, \citep{Ivanovski2014} the motion of rotating aspherical dust grains were studied. Using the observations of the OSIRIS cameras the possible physical parameters of the big grains were evaluated. These first results motivate us to re-evaluate the physical and dynamical properties of cometary dust.} 
 
In this paper we make use of two models for dust particles: the well-established ballistic aggregate model and the hierarchic aggregate model. For each, we calculate dynamic and optical properties such as effective cross-section, extinction coefficient, and drag coefficient. We take into account both solar and cometary forces (gravity, light pressure, gas drag) and examine a wide range of dust particle sizes, from sub-micron to millimeters. We assess the relative importance of the different forces and estimate the characteristic velocities that dust particles acquire near the comet nucleus.  
 
A general description of observable cometary dust properties is presented in Section \ref{ObservableProperties}. In Section \ref{GeometricalProperties}, we briefly report on the geometrical properties of the porous aggregates considered here. Derived dynamical characteristics are given in Section \ref{DynamicalProperties}, where we demonstrate the considerable difference between random porous aggregates and spherical particles. In Section \ref{ForceComparison}, we evaluate the effective forces acting on the different types of aggregates, for a very wide range of dust sizes. Section \ref{InnerComaVelocity} examines dust motion in the innermost coma, i.e. at distances less than 50 km from the nucleus. We discuss our results and compare our model predictions with the first results obtained by instruments on-board Rosetta in Section \ref{Discussion}. Our results were used to simulate the dust-gas inner coma of 67P/Churyumov-Gerasimenko (hereafter, 67P; \citep{Thomas}). Finally, we conclude in Section \ref{Conclusion}.

\section{Observable properties of cometary dust}
\label{ObservableProperties}

Several articles have been published on the dynamical and optical properties of cometary dust. The primary goal of these papers is the analysis and interpretation of various astronomical observations. Below we summarise a few general results relevant here, originating from ground based and in situ observations (obtained from deep space missions), as well as numerical simulations of cometary dust.  
 
It is now well established that cometary dust is very porous and that it possesses low density and strength, as initially proposed by \citet{Whipple}. This result has been validated by numerous ground-based and space observations, and is also supported by current theories of cometary formation \citep{Johansen,Skorov2012,Blum}.
 
Early evidence for the properties of dust was obtained from the analysis of stratospheric interplanetary dust particles. \citet{Bradley} showed that the internal structure of delicate chondritic micrometeorites is highly porous and fragile. Later \citet{Greenberg} used cometary observations to conclude that cometary grains should be extremely porous (porosity $>90$\%). Similar accounts of the high porosity of dust grains were obtained through analysis of photometric observations of the dust coma \citep{Kolokolova2004}.  
 
Recent developments in computer modeling of light scattering by dust particles have shown that ground-based  observation of the dust polarisation curve, the dust colour and the dust albedo are all consistent with comet dust consisting of porous aggregates of sub-micron monomers \citep{Kolokolova2007,Kolokolova2010}. Quantitative studies of polarimetric data have added that the negative branch of linear polarization often seen at small phase angles requires the addition of large porous aggregates into the computer models of cometary dust \citep{Kimura,Kolokolova2007}. The authors further conclude that the dust particles contributing mostly to the bulk light-scattering and thermal-emission properties of comet comae should be larger than 10 $\mu$m. Moreover, \citet{Petrova} showed that a well-defined negative polarization branch was observed if the monomer size parameter ($x_m=2 \pi a_m/\lambda$, the ratio of its characteristic dimension, $a_m$ and the observed wavelength, $\lambda$) was about 1-2, which in the case of optical observations corresponds to monomer sizes about 0.1-0.2 $\mu$m. {\color{red}Hereafter we assume that the size of monomer is 0.1 $\mu$m.}  
 
\citet{Hadamcik} found evidence for large fluffy particles in polarimetric observations of comet 67P during the 2008-2009 apparition. They also detected the negative branch of polarization and interpreted it as a signature of the sub-micron monomer constituents of the larger fluffy aggregates. Spitzer observations of 67P at a heliocentric distance of 5 AU are consistent with the results listed above \citep{Kelley}.  \citet{Fulle} used ground-based and Spitzer observational data of 67P to predict that at Rosetta operation distances the mass of ejected dust would be dominated by grains larger than a few mm. 
 
In summary, the existing observational data strongly suggest that: 1) cometary dust is composed of porous refractive particles, often referred to as fluffy aggregates; 2) these aggregates are made of small sub-micron grains (monomers); 3) the size of these aggregates varies from microns to millimeters.

\begin{table*}[ht]
\centering
\caption{Types and correspondence of ballistic aggregates. BPCA = Ballistic Particle-Cluster Aggregates, BA = Ballistic Aggregates, BAM$m$ = Ballistic Aggregates followed by $m$ Migrations, BCCA = Ballistic Cluster-Cluster Aggregates, [Type]$^n$ = Ballistic Hierarchic Aggregates, where ``Type'' is any of the other aggregate types and $n$ is the number of hierarchic levels. References: \citet{Kozasa}, \citet{Shen}.} \label{BallisticAggregates}
  \begin{tabular}{ccccc} 
  \hline Kozasa et al. & Shen et al. & This work & Fill.\ Factor & Fractal Dim. \\
  \hline 
  BPCA & BA & BA & 0.15 & $\sim$3 \\
  $\cdots$ & BAM1 & BAM1 & 0.20 & $\sim$3 \\
  $\cdots$ & BAM2 & BAM2 & 0.30 & $\sim$3 \\
  BCCA & $\cdots$ & BCCA & $\leqslant$0.03 & $\sim$2 \\
      \hline
  $\cdots$ & $\cdots$ & [Type]$^n$ & (Type fill.\ factor)$^n$ & 2-3 \\
      \hline
  \end{tabular}
\end{table*}

\section{Geometrical properties of aggregates}
\label{GeometricalProperties}

Bearing in mind the purpose of our study, we shall confine our efforts to three interrelated issues: 1) the generation of porous aggregates by numerical methods and the assessment of their geometrical properties, 2) the calculation of optical properties of such aggregates and their interaction with radiation pressure, 3) the calculation of gas-dynamic properties of such aggregates (e.g.\ reaction to gas drag). 
 
\begin{figure*}[ht]
\centering\includegraphics[width=0.6\linewidth]{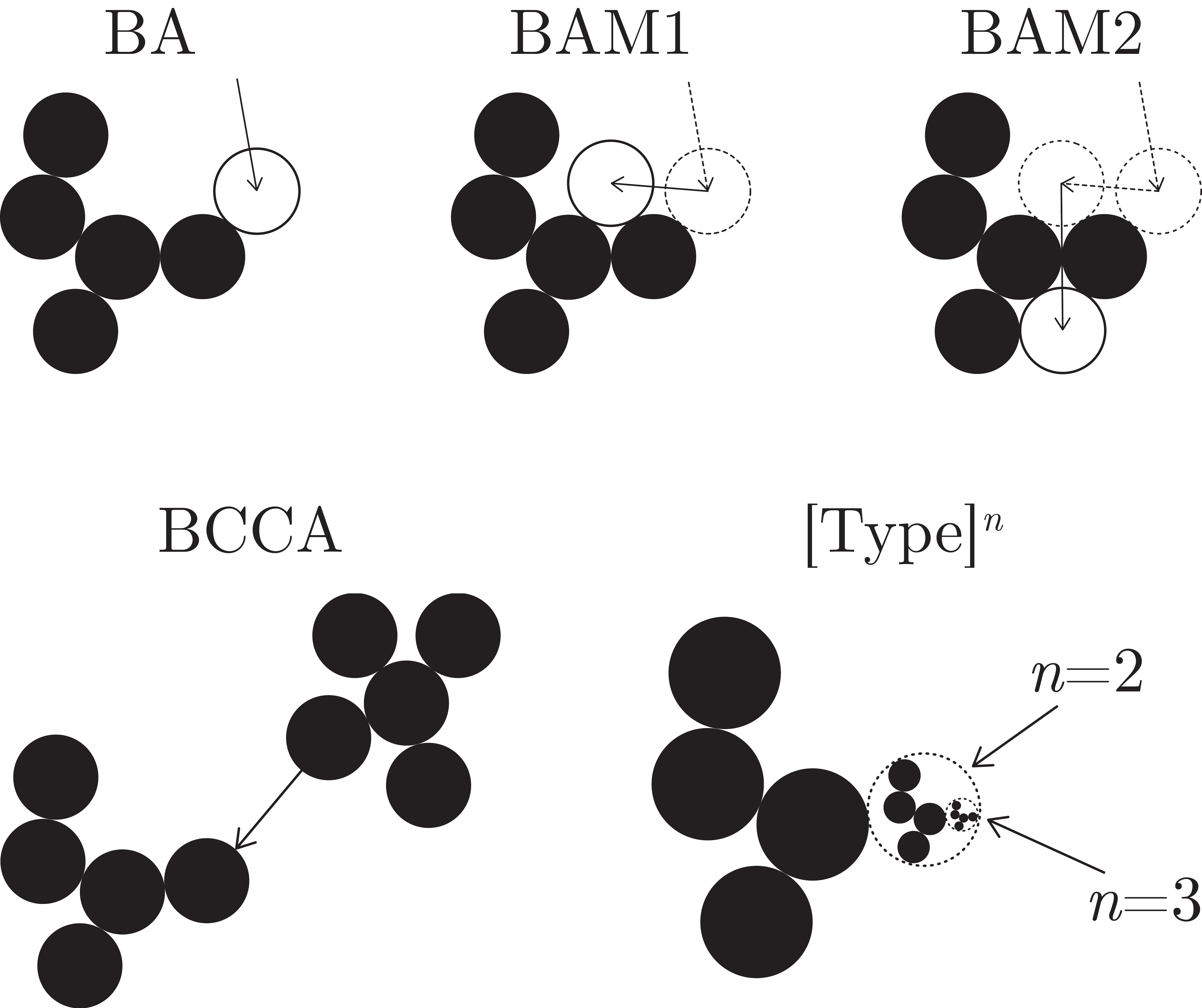}
\caption{Illustration of ballistic aggregates. See Table\ \ref{BallisticAggregates} and description in the text.}
\label{fig:1}
\end{figure*}

We start by describing methods for generating the random porous aggregates meant to simulate interplanetary and cometary dust (see Figure\ \ref{fig:1} and Table\ \ref{BallisticAggregates}). The classical approach is to consider ballistic aggregation of indivisible spherical particle (or monomers) which stick at the first point of contact within a growing cluster (or aggregate). Structures formed in this way, known as ballistic particle-cluster aggregation (BPCA), have the highest filling factor among random aggregates. Such an approach was used by \citet{Kozasa,Kimura,Skorov2008,Skorov2010}.  
 
{\color{red}\citet{Shen} introduced interesting modifications to standard BPCAs: they allowed for monomer migration (by rolling or sliding over the first-contact monomer) producing even more compact, quasi-random, ballistic aggregates. Following \citep{Shen}, we consider three sub-classes to ballistic aggregates. The first, BA, corresponds to classical BPCA and results in the most porous aggregates. The remaining two, BAM1 and BAM2, which allow for one or two migrations respectively, produce aggregates of increasing density. For instance, for clusters with a large number of monomers, $N$, the BA clusters have a filling factor (fraction of volume taken up by monomers) about 0.15, whereas the BAM2 clusters have filling factors of about 0.30, i.e., twice the effective density. We note that all three resulting variants (BA, BAM1, BAM2) have fractal dimension $D_f=3$. See \citep{Shen} for more detail on how these aggregates are generated.}

\begin{figure*}[ht]
\centering\includegraphics[width=0.9\linewidth]{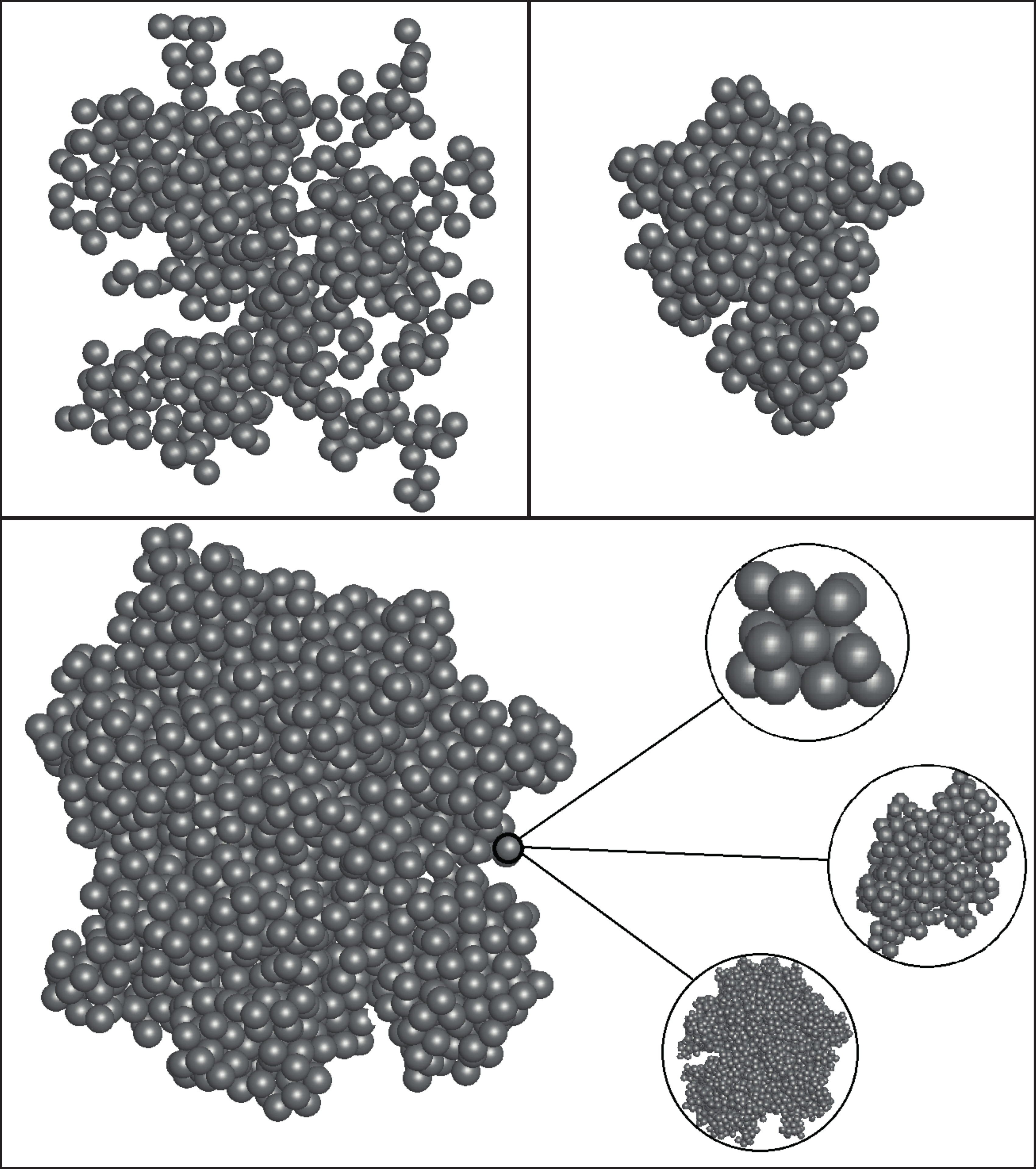}
\caption{Examples of aggregates. Top left = BA {\color{red}(512 monomers)}, top right = BAM2 {\color{red}(512 monomers)}. Bottom = hierarchic BAM2: {\color{red}2048 pseudo-monomers, each pseudo-monomer} is an aggregate composed of a different number of monomers (see description in the text).}
\label{fig:2}
\end{figure*}

In this work, we consider BA, BAM1 and BAM2 aggregates made up of 8 to 4096 single-sized monomers. Examples of BA and BAM2 aggregates consisting of 512 monomers are shown in Figure \ref{fig:2} (top panel). Generally, for all variants, clusters with larger $N$ tend to be more spherical. Typical aggregate asphericity (deviation from unit ratio of minimum and maximum projection area) varies between 20-30\% for BA and 10-15\% for BAM2.  

A different ballistic process is used for generating very porous, open random agglomerates, known as ballistic cluster-cluster aggregation (BCCA; \citep{Kozasa,Mukai1992}). BCCA particles are obtained by random aggregation of equal-mass dust agglomerates. The fractal dimension of the resulting aggregates is close to $D_f=2$, and the filling factor is $\leqslant$3\% for large numbers of monomers (N$\geqslant$256). A detailed comparison of the different algorithms for ballistic agglomeration of dust particles can be found in \citep{Okuzumi}. Hereafter, we consider BCCA as a variant of the most fluffy aggregates. When evaluating the geometrical and dynamical properties of these units, we will use analytical formulae from \citep{Minato2006}.

\begin{figure*}[ht]
\centering\includegraphics[width=0.9\linewidth]{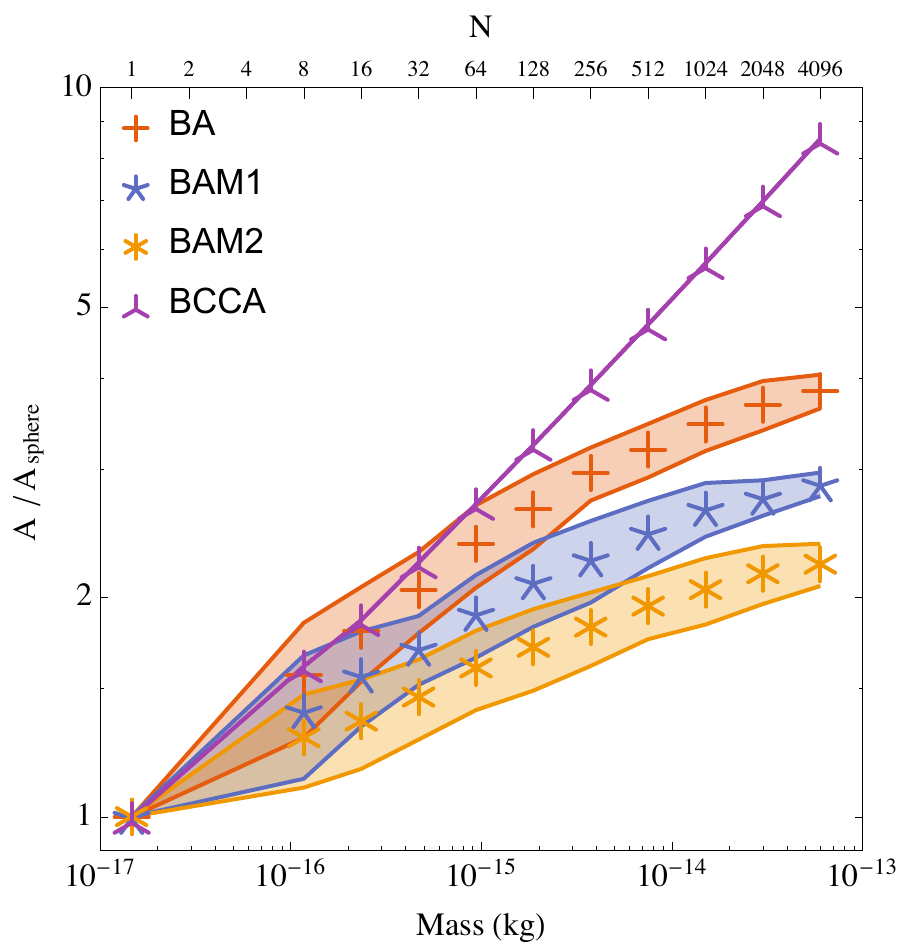}
\caption{Ratio of the cross-section $A$ for BCCA, BA, BAM1, BAM2 clusters made up of $N$ monomers to the cross-section of a sphere $A_{sphere}$ with the same {\color{red}mass}.}
\label{fig:3}
\end{figure*}

We now evaluate the geometrical properties (i.e.\ effective size and cross-section) of all the types of aggregates listed above. The effective cross-section of the aggregates is estimated using a Monte Carlo method. For a given number of monomers $N$, we generate 5 random aggregates of each type. Then, for each random aggregate we calculate three orthogonal projections. The resulting 15 projections are then used to find the average cross-section for each specific type and size of aggregate. In Fig.\ \ref{fig:3} we show the ratio of calculated cross-section of an aggregate {\color{red}$A$} to the cross-section of solid sphere $A_{sphere}$ of the same mass. We present results for the three types of particle-cluster aggregates (BA, BAM1, BAM2) and for the cluster-cluster aggregates (BCCA). For the largest aggregates ($N=4096$) with fractal dimension $D_f=3$ the ratio varies between 2 (BAM2) and 4 (BA). As BCCAs have fractal dimension $D_f=2$, their cross-section is proportional to mass. Thus, the ratio of BCCA cross-section to the cross-section of solid sphere is a linear function. For $N=4096$ this ratio is about 9. To convert number of monomers to aggregate mass we assume hereafter that bulk density of material is 2400 kg m$^3$ \citep{Blum}.

\begin{figure*}[ht]
\centering\includegraphics[width=0.9\linewidth]{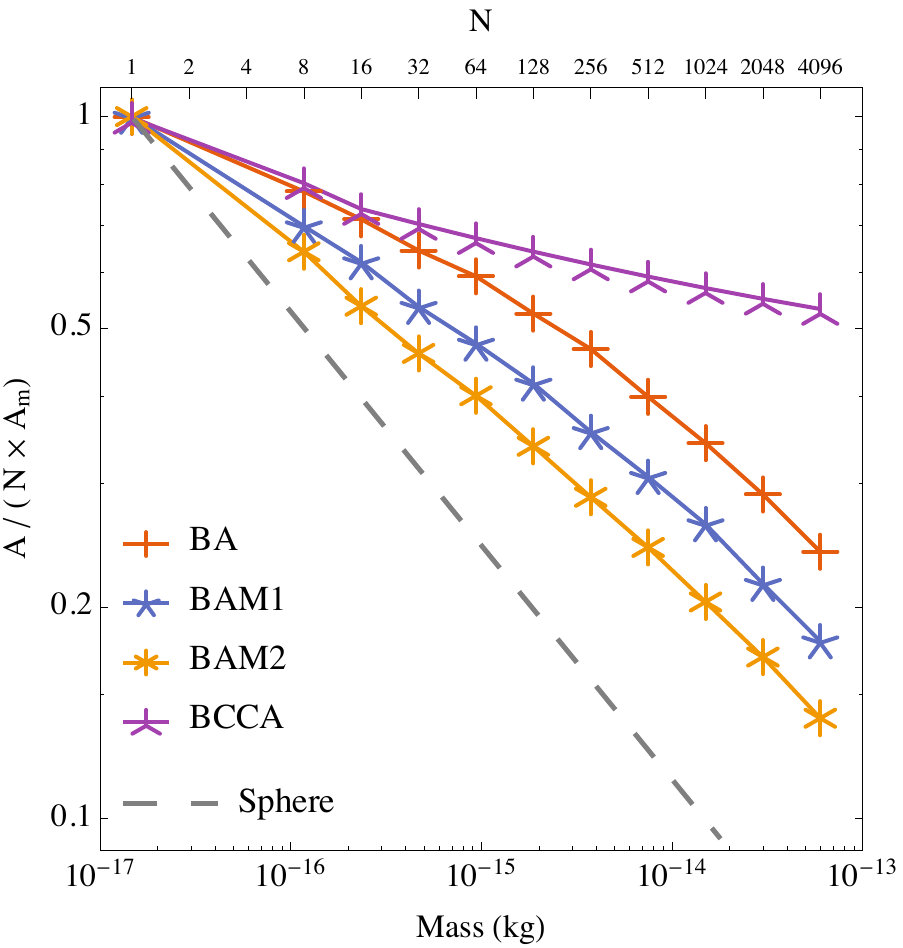}
\caption{Shadowing effect for different types of aggregate (BCCA, BA, BAM1, BAM2): Ratio of aggregate cross-section $A$ to total cross-section of constituent monomers $N\times A_m$.}
\label{fig:4}
\end{figure*}

Another interesting property describing the degree of compactness of an aggregate is the ratio of its cross section {\color{red}$A$} to the total cross-section of all its constituent monomers $N\times A_m$. This value, which characterizes the degree of shadowing, is shown in Fig.\ \ref{fig:4} for all types of aggregates. For the open fluffy BCCA units the shadowing effect is relatively small: even for the largest aggregates more than 50\% of total cross-section of monomers is still visible. Instead, in the case of the largest BA and BAM2 aggregates the shadowing increases and only respectively 25\% and 15\% of the monomer cross-section is visible. 
 
As noted in the Introduction, observations of the negative branch of the polarization curve suggest that the size of the monomers in our aggregates should be about 0.1-0.2 microns. Hence, our largest aggregates are representative of dust particles only a few microns in size. However, the first results obtained by instruments on-board Rosetta spacecraft \citep{Rotundi,Schulz} indicate that the nucleus ejects porous particles hundreds of microns in size at $\sim$3.6 AU. To study the dynamic properties of such large aggregate we consider two approaches. First, we use a standard extrapolation procedure to calculate the geometric properties of large aggregates having the same structure as the smaller ones simulated above. We extrapolate the ratio of cross-sections $A/A_{sphere}$ as a logarithmic function of the monomer number $N$. {\color{red} Note that all comparisons shown below are made for the case when the bulk density of the monomers is equal to the bulk density of a solid sphere.} Only aggregates containing more than 16 monomers are used for extrapolation procedure.  
 
Second, we produce large porous aggregates using a hierarchic procedure \citep{Skorov2012} that generates aggregates of aggregates (see Figure \ref{fig:1} and Table \ref{BallisticAggregates}). This approach can be considered as a variant of the classical BCCA method. {\color{red}However, there is an important difference, namely: the BCCA is created due to recursive collisions of two aggregates constructed in the previous step, whereas the hierarchic aggregate is created from pseudo monomers of the same size}. Similar approaches were used by \citet{Shkuratov} and \citet{Okuzumi}. {\color{blue}
Each pseudo-monomer is an aggregate constructed at the previous step 
of simulation. That is, the first level hierarchical aggregate is built
from solid spherical monomers, second level aggregates are built from
first-level aggregates (BPCAs) which are considered now as pseudo-monomers. At the third level, second-level aggregates  play the role of pseudo-monomers, etc.  An example of such a second level hierarchical aggregate is shown in Figure \ref{fig:2} (bottom panel)}. 
 
We construct hierarchic aggregates of the second level using BA, BAM1 and BAM2 aggregates as building blocks (i.e.\ new pseudo-monomers). For a number of hierarchic levels $n$ we refer to a hierarchic aggregate of a give type by [Type]$^n$. First we determine the size of each building block, $N_m$ (this parameter may vary from 8 to 4096 monomers), and then we determine the number of building blocks, $N_a$, in the constructed aggregate. {\color{red} Note that $N_m$ is constant for a given aggregate.} Thus, an aggregate containing the same number of real monomers can be constructed in different ways (varying both $N_m$ and $N_a$). During aggregation, the effective {\color{red}(mean)} size of the pseudo-monomers composing our hierarchic aggregate is recalculated. To evaluate this we simulate random ballistic collisions between two aggregates of the same size and determine the effective distance between their centers of mass. Half the distance is taken to be the effective size of the pseudo-monomers. Generally, a few thousand binary collisions are needed to get a statistically reliable size. Because the largest ballistic aggregate contains 4096 monomers, the largest hierarchic aggregate contains $4096\times 4096$ monomers, and only one realization of it in terms of $N_m$ and $N_a$ is possible in our model. For all smaller [Type]$^n$ aggregates there are several possible variants of clusters with the same number of monomers.  A [BAM2]$^2$ aggregate comprising 2048 pseudo-monomers is shown in Figure \ref{fig:2} (bottom panel). Each pseudo-monomer may be a ballistic aggregate of different size: the insets in the Figure show the different BAM2 aggregates used as pseudo-monomers. Note that the size of a pseudo-monomer is usually smaller than the radius of the circumscribed sphere (shown in the Figure) due to the random averaging. By applying this technique we are able to build aggregates with sizes up to hundred microns.

\begin{figure*}[ht]
\centering\includegraphics[width=0.9\linewidth]{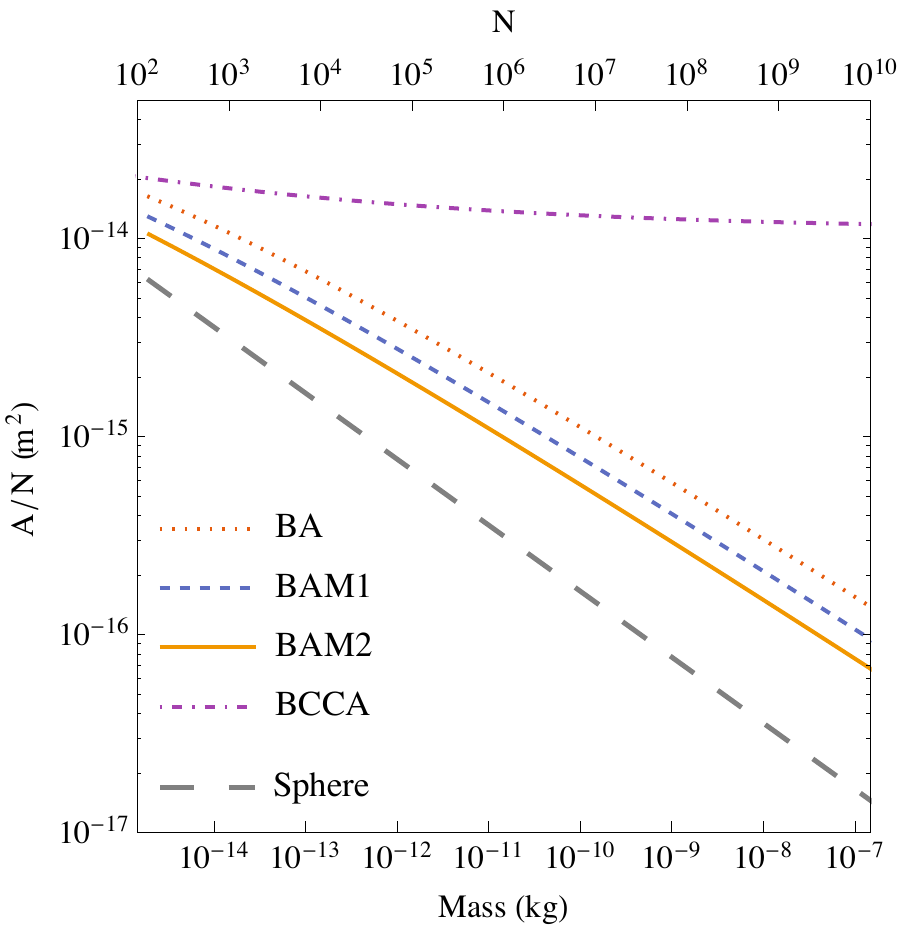}
\caption{Extrapolated specific cross-section (ratio of aggregate cross-section, $A$ to number of monomers, $N$) as a function of the number of constituent monomers for BCCA, BA, BAM1, and BAM2 aggregates.}
\label{fig:5}
\end{figure*}

It is interesting to compare the results obtained using the two approaches just described. Figure \ref{fig:5} shows the ratio of extrapolated agglomerate cross-section to the number of monomers as a function of the number of monomers for all types of ballistic aggregates and for solid spheres of equivalent mass. Because aggregates of different types have different fractal dimensions their approximate sizes are also different for a given number of monomers. However, using the relation $N \sim r^D_f$ \citep{Meakin1987}, where $r$ is so called radius of gyration, we can roughly evaluate the maximum size of aggregates presented in the Figure: cluster-cluster aggregates have a gyration radius about 1 mm, particle-cluster aggregates have a gyration radius about 100 $\mu$m, and solid spheres have sizes roughly 3-10 times smaller than the size of particle-cluster aggregates (i.e., a few tens of $\mu$m). Strictly speaking, the filling factor of fractal aggregates with $D_f=2$ should be constant when its mass goes to infinity, which means that curves for particle-cluster aggregates should be parallel to the curve for solid spheres. However, our extrapolation based on the small aggregates ($N < 4096$) that we are able to simulate yields slightly fluffier, homogeneous aggregates. For our purposes, this small difference is unimportant.  

\begin{table}[ht]
\centering
\caption{Ratio of aggregate to sphere cross-section as a function of particle mass for hierarchical aggregates of type [BA]$^2$ and [BAM2]$^2$. See Figure \ref{fig:6}.} \label{AoverAsphere}
  \begin{tabular}{ccc} 
  \hline   
  \noalign{\smallskip}
  $\log_{10}$Mass & [BA]$^2$ & [BAM2]$^2$ \\
  (kg) & $A/A_{sphere}$ & $A/A_{sphere}$ \\
  \noalign{\smallskip}
  \hline 
  \noalign{\smallskip}
$-16$ & $1.88  \pm  0.32$ & $1.36 \pm 0.08$ \\
$-15$ & $2.93  \pm  0.38$ & $1.70 \pm 0.11$ \\
$-14$ & $4.66  \pm  0.82$ & $2.20 \pm 0.18$ \\
$-13$ & $6.94  \pm  1.07$ & $2.78 \pm 0.18$ \\
$-12$ & $9.32  \pm  1.42$ & $3.37 \pm 0.24$ \\
$-11$ & $12.89 \pm  1.22$ & $4.22 \pm 0.31$ \\
$-10$ & $15.73 \pm  0.62$ & $5.03 \pm 0.16$ \\
  \noalign{\smallskip}
  \hline
  \end{tabular}
\end{table}

\begin{figure*}[ht]
\centering\includegraphics[width=0.9\linewidth]{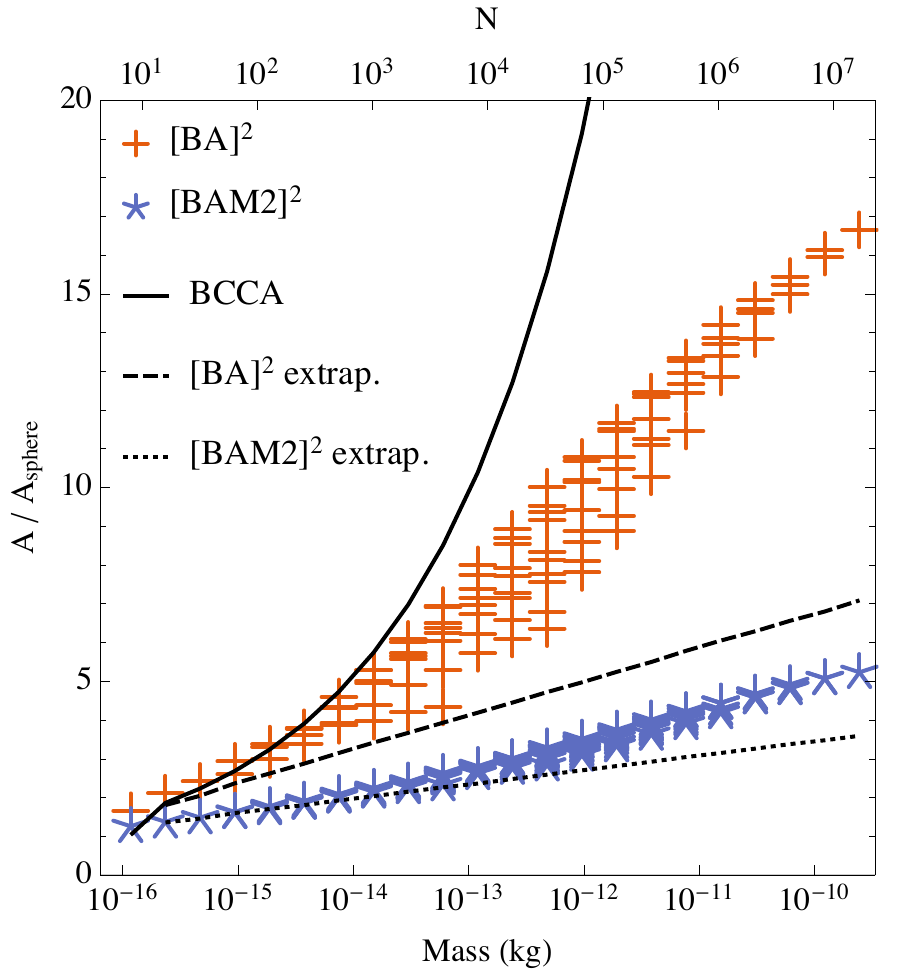}
\caption{Ratio of aggregate cross-section $A$ to cross-section of an equivalent mass sphere $A_{sphere}$ for BCCA and hierarchic aggregates [BA]$^2$ (crosses) and  [BAM2]$^2$  (stars). Lines ({\color{red}dashed and dotted}) show extrapolated relations for [BA]$^2$ and [BAM2]$^2$ respectively. See also Table \ref{AoverAsphere}.}
\label{fig:6}
\end{figure*}

Figure \ref{fig:6} shows the ratio of cross-section of large aggregates to the cross-section of solid spheres of the same mass. Two types of aggregates having the same fractal dimension $D_f=3$ are compared: the densest (BAM2) and the fluffiest (BA) aggregates. This Figure shows results obtained by extrapolation for the case of homogeneous aggregates (solid lines) and by explicit calculation for the case of hierarchic aggregates (crosses and triangles). Because for fixed number of monomers (i.e., fixed mass) there are several ways to construct a [Type]$^n$ hierarchic aggregate (see explanation above) the total cross-section depends on the size of the constituent aggregates. For example, a [BA]$^2$ cluster consisting of a total $N=8194$ monomers can be built from $N_a=2$ ballistic aggregates, each with $N_m=4096$ monomers, or from $N_a=128$ aggregates, each with $N_m=64$ monomers. Obviously, the dynamic properties of the two might be different. For a mass of about $10^{-13}$ kg the range of cross-sections is about a factor 2 for hierarchic aggregates of type [BA]$^n$, but only 20\% for aggregates of type [BAM2]$^n$. This is why we find considerable variations in total cross-section for the same mass. The implication is that the total cross-section of an aggregate is not entirely defined by its fractal dimension and total number of monomers. The largest [Type]$^2$ aggregates are composed of 4096 pseudo-monomers, each of which in turn consists of 4096 real monomers. The size of these hierarchic aggregates reaches hundreds of microns depending on the type of aggregate. One can see that the cross-section values obtained by extrapolation are {\color{blue} half} for the largest aggregates, i.e., our hierarchic aggregates are fluffier and have smaller filling factor than similar homogeneous aggregates. This is not surprising, since the average void size in the [Type]$^2$ aggregates is larger. Thus, the total cross-section of our [Type]$^2$ aggregates are widely scattered between the curve for BCCA aggregates (filling factor 2\%) and the curve for densest homogeneous BAM2 aggregates (filling factor 30\%). Table \ref{AoverAsphere} lists the mean ratio of aggregate to sphere cross-section as a function of aggregate mass.

\section{Dynamical properties of aggregates}
\label{DynamicalProperties}

The different porous aggregate models described above serve as the basis for the study of the optical and dynamic properties of interplanetary and cometary dust. In this section we will focus on the effects of radiation pressure, gas drag and gravity acting on dust grains in the vicinity of a cometary nucleus.  

\subsection{Radiation pressure, $F_R$}

Light scattering by irregular porous aggregates has been numerically simulated using a number of different approaches, such as: 1) the discrete dipole approximation (DDA; \citep{Draine}), 2) the superposition T-matrix method \citep{Mishchenko}, 3) simple Mie theory \citep{vdHulst}, 4) Mie theory combined with the Maxwell-Garnett mixing rule (MG-Mie; \citep{Bohren,Mukai1992}), and 5) the geometrical optics approximation (GO; \citep{Shkuratov}). Methods 1) and 2) are time consuming and require considerable computing power, so they are generally only applied to aggregates that are comparable to the radiation wavelength. When dealing with large dust grains, methods 3) to 5) are more commonly used.  
 
Here, we estimate the effects of radiation pressure on dust aggregates using the formalism outlined in \citep{Mukai1992} and \citep{Okuzumi}. We calculate the geometrical scattering cross-section, $A$, and the efficiency factor, $Q_{PR}$, which depends on grain size, structure and composition. 

The radiation pressure force, $F_R$, acting on a dust particle of effective cross-section $A$ at heliocentric distance $R$ is given by: 

\begin{equation}
\label{eq:FR}
F_R=\left(\frac{A}{c}\right)\left(\frac{R_0}{R}\right)^2 \int_0^\infty B_0\left(\lambda\right)Q_{PR}^*\left(m^*,\lambda\right)\textrm{d}\lambda
\end{equation}

\noindent where $c$ is the speed of light, $R_0$ the solar radius, $B_0$ the radiance of the sun at wavelength $\lambda$, and $Q^*_{PR}$ the radiation pressure efficiency coefficient, which depends on $\lambda$, on the refractive index $m^*$, and on the grain shape and spatial structure. For the $B_0$ function we use data from the National Renewable Energy Laboratory\footnote{http://rredc.nrel.gov/solar/spectra/am0/}.  

Following \citep{Burns} we calculate the radiation pressure efficiency as $Q^*_{PR} = Q_a + Q_s\left(1-\left<\cos\lambda\right>\right)$, where $Q_a\left(m^*,\lambda\right)$ is the effective absorption cross-section, $Q_s\left(m^*,\lambda\right)$ is the effective scattering cross-section, and $\left<\cos\lambda\right>$ is the average cosine of the scattering angle characterizing the asymmetry of the scattering function of the grain. As mentioned above, the DDA method can be applied to small aggregates, up to a few thousand of monomers. An application can be found, for example, in \citep{Kohler} and \citep{Shen}. Note that those simulations clearly demonstrate a strong dependence of both scattering and absorption cross-sections on grain composition. Numerous studies show that cometary dust has a complex chemical composition, including organic components \citep{Kissel,Schulz}, and one can expect that the main source of uncertainty in dust optical properties is due to the uncertain composition. An accurate determination of the effective refractive index is beyond the scope of this paper, so we adopt two simplifying approaches: 1) we consider two common materials, namely astronomical silicates and graphite, and 2) we use Mie theory for spheres. The latter approximation is justified by the results of \citet{Mukai1992} and \citet{Mukai2005}. It is well established (e.g., \citep{Kokhanovsky}) that the extinction coefficient increases dramatically for grain sizes comparable to the wavelength of radiation (resonance region) and that it rises to the value 2 for very large grains independently of grain composition and shape (extinction paradox).  

Even when simple dust species are modeled, significant uncertainty in their properties remains. For example, while many papers devoted to the optical properties of silicates refer to the refractive index published in \citep{Kozasa} and use a density for silicate material of 2400 kg m$^{-3}$, \citet{Laor} suggest a significantly different refractive index and a higher density (3500 kg m$^{-3}$) for the same material. The particular choice of density is important in the current context because it impacts the parameter $\beta$, the ratio of radiation pressure to solar gravity. Our tests show that for the case of silicate $\beta$ can vary considerably for the same aggregate if different sources are used. 
In order to avoid this uncertainty we consistently use the data presented by \citet{Laor}. They calculate the absorption and scattering efficiency coefficients and average the asymmetry factor and the refractive index $m^*$ for a wide range of grain sizes using different techniques (Rayleigh-Gans theory, Mie theory) as a function of wavelength.

\subsection{Gas drag force, $F_{gd}$}

The main process releasing cometary dust from the nucleus is the sublimation of its volatile components. Hence, we investigate the gas drag force $F_{gd}$ acting on irregular porous grains embedded within the expanding gas flow. Note that for the dust fraction cometary gas is collisionless, i.e.\ molecular mean free path is much larger than the dust size even at small heliocentric distances. In this regime of dust-gas interaction the gas drag force $F_{gd}$ depends on many parameters (see for details, \citep{Bird,Skorov1999}) and is usually expressed as:  

{\color{red}
\begin{equation}
\label{eq:Fgd}
F_{gd}\left(l\right)=\frac{1}{2}\,C_D\, A\,n_l\, m_{\mathrm{H}_2\mathrm{O}} \left(v_l-v_d\right)^2,
\end{equation}
}

\noindent where $C_D$ is the drag coefficient, $m_{\mathrm{H}_2\mathrm{O}}$ is the {\color{red}water} molecule mass, $A_d$ is the effective particle cross-section, $n_l$ and $v_l$ are the gas density and velocity, $v_d$ is the dust particle velocity. The effective cross-section of fluffy aggregates $A$ is discussed in detail above. The gas number density $n_l$ is evaluated assuming gas expansion at constant velocity $v_l$ from a homogeneous isothermal spherical nucleus. In this case we assume that the nucleus radius is 2 km, the surface temperature is $T_s$, and the corresponding gas velocity, $v_l$, is determined by the energy balance at the specific heliocentric distance. We also assume that the total gas production is 2 kg s$^{-1}$ \citep{Gulkis2015}. The efficiency of momentum exchange between gas molecules and the dust grain is characterized by the drag coefficient $C_D$. {\color{red} In general, one should take into account that the cometary coma contains various gas species, which have different masses and velocities. The presented equation is valid for the case when water is the dominant species.} 
 
If the grain size is much smaller than the gas molecule mean free path $l$, the macroscopic velocity of gas $v_l$ is much smaller than thermal velocity $v_T$, and the gas is in kinetic equilibrium (i.e.\ the distribution function of molecule velocities is Maxwellian), the drag coefficient on a sphere is about $C_D=8/3$ \citep{Armitage}. For specular scattering, the drag coefficient does not depend on the temperatures of gas and dust and approaches $C_D=2$ as the ratio $v_l/v_T$ goes to infinity. In cometary physics both simplifications are invalid. In the general case for spherical particles, $C_D$ is expressed by a more complicated formula \citep{Baines}:  

\begin{multline}
\label{eq:CD}
C_D =\frac{2W^2+1}{\sqrt{\pi}W^3}\exp\left(-W^2\right) +\frac{4W^4+4W^2-1}{2W^4}\mathrm{erf}\left(W\right) \\ +\frac{2\sqrt{\pi}\left(1-\epsilon\right)}{3W}\sqrt{\frac{T_d}{T_l}}
\end{multline} 

\noindent where W is the speed ratio ($W=v_l \sqrt{m_{\mathrm{H}_2\mathrm{O}}/2kT}$), $T_d$ and $T_l$ are the dust and gas temperatures, and $\epsilon$ is the specularly reflected fraction of molecules. 
 
The microscopic structure of grains (i.e., porosity and fractality) should be taken into account when this formula is applied for aggregates considered here. The drag coefficient for fluffy ballistic aggregates was investigated by \citet{Meakin1989}, \citet{Nakamura} and \citet{Blum1996}. They treated gas molecules as point masses and calculated collisions and  reflection using a Monte Carlo technique. Because the drag coefficient is proportional to the impulse transfer per collision, they compared the effectiveness of collisions for solid spheres, for BPCA, and for BCCA aggregates, and found that the mean momentum transfer effectiveness is about 10\% larger for BCCA and 20\% larger for BPCA in the case of specular reflection. \citet{Meakin1989} found that the effectiveness of momentum transfer for diffuse reflection is about 30\% larger compared to the sphere. Note also that these excesses decrease with increasing aggregate size. However, the relative increase of momentum transfer efficiency plays only a minor role when compared to the relative increase of cross-section $A$ for fluffy grains of the same mass. Thus, we can conclude that Eq.(\ref{eq:Fgd}) is accurate enough for calculating the gas drag force on porous irregular aggregates if the correct ratio between dust cross-section and mass is applied.   
 
We point out two caveats of our approach. First, we assume that velocity distribution function is Maxwellian, which is likely unwarranted for ice sublimating into vacuum. The kinetic Knudsen layer is adjacent to the surface of the cometary nucleus: here the initially non-equilibrium velocity distribution function of gas molecules relaxes to the Maxwell equilibrium distribution function and, as a result, the macro-characteristics of the gas flow vary several-fold. This effect is well known in cometary physics (\citep{Shulman}, \citep{Crifo}). \citet{Skorov1999} calculated the drag force acting on a sphere and a cylindrical disk for an arbitrary combination of diffuse and specular scattering for a shallow plane layer. However, the assumption that the non-equilibrium layer is shallow is obviously invalid when the number density of the gas is low and the sublimating surface is not an infinite plane. If the total gas production is 2 kg s$^{-1}$, the mean free path for water molecules is hundreds of meters, i.e., comparable with the nucleus radius. In this case, due to the spherical expansion, gas equilibrium is not achieved. A second caveat deals with the diffuse scattering by dust: in this case, $C_D$ depends on the dust temperature $T_d$, which should be calculated from the general radiative energy balance. But the temperature of small dust grains might be much higher than gas temperature \citep{Lien}.  Thus, the drag coefficient should be recalculated for rarefied gases and hot dust grains. This will be addressed in a future paper.

\subsection{Gravity forces}

The dust particles lifted from the comet surface are also subject to gravity due to the Sun and the comet nucleus. The relative effect on dust grains of solar radiation pressure and solar gravity are often combined in a single parameter, $\beta=F_R/F_{GS}$, where $F_R$ is the force due to radiation pressure and $F_{GS}$ is the force due to solar gravity. Both forces vary with the inverse square of the heliocentric distance, so $\beta$ is a dimensionless parameter quantifying the relative importance of radiation pressure in the dynamical evolution of dust particles. Because we are interested in the motion of cometary grains near the comet nucleus, we introduce a new parameter, $\beta_c$, which measures the relative importance of solar radiation pressure and gravity due to the nucleus, $F_{GC}$. This new parameter is a function of both the heliocentric distance and the intensity of the cometary gravity field. We assume that the comet nucleus is a homogeneous sphere with radius $R_C=2$ km and average density $\rho=470$ kg m$^{-3}$ \citep{Sierks}. It is interesting to note that at a heliocentric distance considered here, $R_H=3.2$ AU, solar gravity is about three times stronger than cometary gravity even at the surface of a 2 km sphere.

\section{Force Comparison}
\label{ForceComparison}

The forces acting on the dust aggregates just above the nucleus surface are compared in Fig.\ \ref{fig:7}. We show ratios for various pairs ($F_{gd}/F_{GC}$, $F_{gd}/F_R$ and $F_R/F_{GC}$) and for extreme variants of aggregates (BA and BAM2). These ratios are plotted as functions of aggregate mass $M$ and total monomer numbers $N$. As expected, the gas drag force $F_{gd}$ dominates near the surface. Solar radiation pressure $F_R$ is visibly higher than cometary gravity $F_{GC}$ for the masses below $10^{-13}$ kg, but becomes only 15\% (for BAM2 aggregates) and 50\% (for BA aggregates) of the gravity force for the most massive particles ($>10^{-10}$ kg). Because the gas drag force is proportional to aggregate cross-section and the gravity force is proportional to the aggregate mass (or monomer number) the ratio $F_{gd}/F_{GC}$ follows the behavior of cross-section to mass ratio (see Fig.\ 4), i.e., it decreases dramatically with size. However, even for the largest and densest aggregates the gas drag force exceeds comet gravity by about twenty times, i.e.\ excluding cohesion, the upper limit of the size of lifted particles is much higher than our aggregates (see \citep{Blum}). Gas drag and radiation pressure are both proportional to the aggregate cross-section $A$ and are independent of aggregate mass $M$, so the variation of the ratio $F_{gd}/F_R$ is due to variation of $Q_{PR}$ as discussed above. This parameter is maximum for grain sizes comparable with the wavelength of radiation, which explains the relatively low ratio for small aggregates ($M\sim 10^{-16}$ kg). The rise towards the lowest mass, which corresponds to a single solid monomer 0.1 $\mu m$, is due to a drop in scattering efficiency. For large aggregates the ratio is doubled as $Q_{PR}$ tends to 2.  

The same forces are compared at ten cometary radii in Fig.\ \ref{fig:8}. Because we assume a constant gas velocity, the forces due to gas drag and comet gravity scale in the same way. Thus, the curves for $F_{gd}/F_{GC}$ are independent of distance from the comet. Both the gas density and gravity decrease hundred times and as a result the radiation force becomes much larger than cometary gravity; even for the densest and largest aggregates (BAM2) the ratio is above ten. At the same time the gas drag force becomes less important and already at this distance from the nucleus it is only twice as large as radiation pressure. Thus, we can conclude that acceleration by gas is the dominant force acting on porous particles smaller than a millimeter near the nucleus surface, at the considered heliocentric distance $R_H=3.2$ AU and for the assumed total gas production 2 kg s$^{-1}$. Clearly, if activity is concentrated in small regions on the nucleus surface this conclusion becomes even stronger. At the same time, at distances of tens of nucleus sizes the radiative pressure becomes dominant, mostly as a result of the spherical expansion of the sublimating gas.

\section{Dust Velocity in the Inner Coma}
\label{InnerComaVelocity}
\begin{figure*}[ht]
\centering\includegraphics[width=0.9\linewidth]{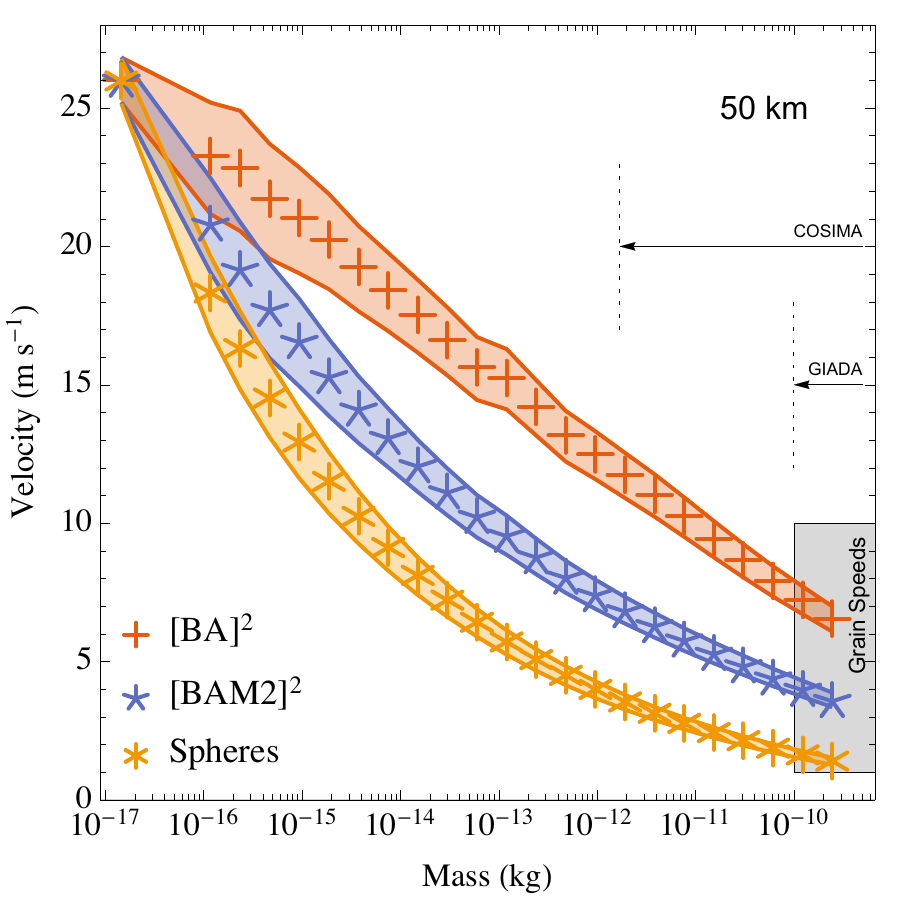}
\caption{Velocity of grains as a function of aggregate mass (and monomer number) at 50 km from the center of mass of the nucleus. Results for BA-, BAM2-type hierarchical aggregates and solid spheres are shown. Lines with symbols correspond to drag acceleration only; bracketing lines indicate variation from anti-solar point to sub-solar point. The COSIMA and GIADA detectability ranges are also indicated.}
\label{fig:9}
\end{figure*}

The effective forces calculated above can be used to evaluate the dust velocity near the comet nucleus. This issue is directly connected with observations by the GIADA instrument on-board Rosetta. In this section, we obtain velocity estimates for a range of particle sizes. The spatial distribution of the smallest particles has important implications for the photometric model of then inner coma, whereas the results for large fluffy particles is useful for interpreting data from the GIADA and COSIMA instruments \citep{Rotundi,Schulz}. Our results are summarized in Fig.\ \ref{fig:9}. Three model geometries are shown: a) dust starts at the sub-solar point (radiation force and comet gravity are opposite to gas drag force), b) dust starts at the anti-solar point (radiation force and gas drag force are opposite to comet gravity); c) dust is accelerated by gas drag force only. The dust velocities reached at a distance of 50 km from the nucleus center are plotted for solid spherical grain, the fluffiest (BA) and the densest (BAM2) model aggregates. As before, we assumed that gas velocity is constant and the nucleus surface is isothermal. {\color{red} The values obtained under these model assumptions provide us a lower limit only. A localised outgassing (i.e.\ powerful collimated jets) would increase the dust speeds.} For the largest particles, the dynamic and optical properties of hierarchic aggregates [Type]$^2$ were applied.

Because the importance of radiation decreases with increasing grain size all curves in Fig.\ \ref{fig:9} converge, and velocity variations for different starting points are unimportant even for micron size dust particles. 
Solid spheres have the smallest corresponding velocity for all range of sizes. As the cross-section-to-mass ratio of solid spheres decreases fastest with mass, the range of velocities is largest for those particles in the mass range considered. One should expect that the velocity range is smallest for hierarchic aggregates with fractal dimension about 2, because in this case the cross-section to mass ratio is nearly constant and, hence, the velocity is independent of mass. The decrease in effective density (or the increase in effective size for a fixed mass) always leads to an increase in dust velocity: for the largest particles ($m=10^{-10}$ kg) the velocity increases from 3 m s$^{-1}$ to 8 m s $^{-1}$. This increase should be highest for the hierarchic aggregates as in this case the velocity of largest particles is about ten times larger than for largest solid sphere and about 25 m s$^{-1}$.

When the velocity is evaluated as a function of distance from the center of mass, we find that all grains achieved about 90\% of their terminal velocity at a distance of about 10 nucleus radii. This value is insensitive to particle size, because it is mainly determined by a dramatic drop in the gas density: for the spherical expansion it drops by two orders of magnitude. It is also interesting to compare the time required by different dust particle sizes to reach the same distance from nucleus. When the mass varies from $10^{-16}$ kg to $10^{-10}$ kg for the BA grains the time increases by only a factor of 4: the small particles reach 50 km after approximately 30 minutes, whereas the largest grains require about 2 hours. In any case the travel time is much shorter than the rotation period of nucleus: in the simplest model of gas activity the difference in the starting points may achieve about 30 degrees.

\section{Discussion and open questions }
\label{Discussion}

\subsection{Geometrical properties of aggregates}

In this paper, we have studied the geometric and dynamic properties of aggregates with direct application to the physics of cometary dust. We have shown that the properties of dust aggregates are a strong function of the method of aggregation. It is traditionally assumed that dust aggregation depends mainly on the macroscopic environmental conditions. We show here that even if we consider only the classical mechanism of ballistic particle cluster aggregation, small modifications of the sticking conditions can lead to dramatically different outcomes.

Following \citep{Shen} we examined ballistic aggregates with markedly different filling factors: from 0.15 for classical BPCA (termed BA in the article) to 0.30 for BAM2. Thus we can expect that even under similar conditions of formation, the effective particle density and cross-section for a fixed dust aggregate mass might vary significantly. The fractal dimension of these aggregates, however, remains the same: $D_f \sim3$. 
 
For the first time, we have thoroughly reviewed hierarchical particles ([Type]$^2$; see Figs.\ \ref{fig:1} and \ref{fig:2}). These compound particles are constructed in a similar way to the well-known ballistic cluster-cluster aggregates (BCCA) as each particle is composed of pseudo-monomers each of which is itself a ballistic aggregate. Because one can use different combinations of number of monomers per aggregate and number of aggregates per dust particle to build [Type]$^2$ of a given mass, the normalized cross-section and effective density of such particles might also vary considerably. Thus, the density of [Type]$^2$ particles of mass about $10^{-13}$ kg can vary by factor 2 (Fig.\ \ref{fig:6}). It is interesting to note that our hierarchical particles are closer to particle-cluster aggregates than to cluster-cluster aggregates, meaning that their fractal dimension is close to 3. Nevertheless, the properties of these particles differ significantly from the corresponding properties of simple ballistic aggregates of the same mass. Systematic simulations show that they can not be obtained just by extrapolation. The largest hierarchic aggregates considered here have cross-sections about $3\times$ larger than ballistic aggregates of the same mass.
 
\subsection{Dynamical properties of aggregates}
 
\subsubsection{Radiation pressure $F_R$}

We have shown that the effect of radiation pressure on dust aggregates is sensitive to the chemical composition of the particles. For small particles ($\sim 0.1$ $\mu$m), the ratio of the radiation pressure on graphite and silicate aggregates may surpass one order of magnitude. For sizes larger than 1 $\mu$m, this ratio is approximately constant and close to 2. In our model the pressure is directly proportional to the effective cross-section of the aggregates. Therefore, the effectiveness of radiation pressure can be characterized by a shadowing coefficient -- the ratio of the aggregate cross-section to the total cross-section of all of its constituent monomers. For BCCA ($D_f \sim2$) this ratio decreases slowly with increasing number of monomers, whereas for the largest BA particles ($N=4096$) this ratio is only $\sim 0.2$, i.e., only this fraction of monomers is directly illuminated. The shadowing effect may be important for the thermal modeling of aggregates containing ice.

\subsubsection{Gas drag force $F_\mathrm{gd}$}

Gas drag force is also proportional to the effective cross-section of an aggregate. Although we used a drag coefficient $C_D$ for spheres, we expect that the influence of the microscopic structure of particles (porosity and fractal dimension) is insignificant. Momentum transfer efficiency is higher for porous aggregates, but only by at most 30\%. The most important model simplification is the assumption of specular scattering of gas molecules. We exclude diffuse scattering in order to eliminate the need to consider the dust temperature in our model. An account of the interaction between the warm, sunlit dust particles and the cold gas flow is beyond the scope of the current paper, as is generalizing the gas molecules velocity distribution beyond Maxwellian equilibrium. The latter will be important in cases of weak activity and gas production from small active regions. We plan to address these issues in a future paper.

\subsubsection{Force comparison}

\begin{figure*}[ht]
\centering\includegraphics[width=0.9\linewidth]{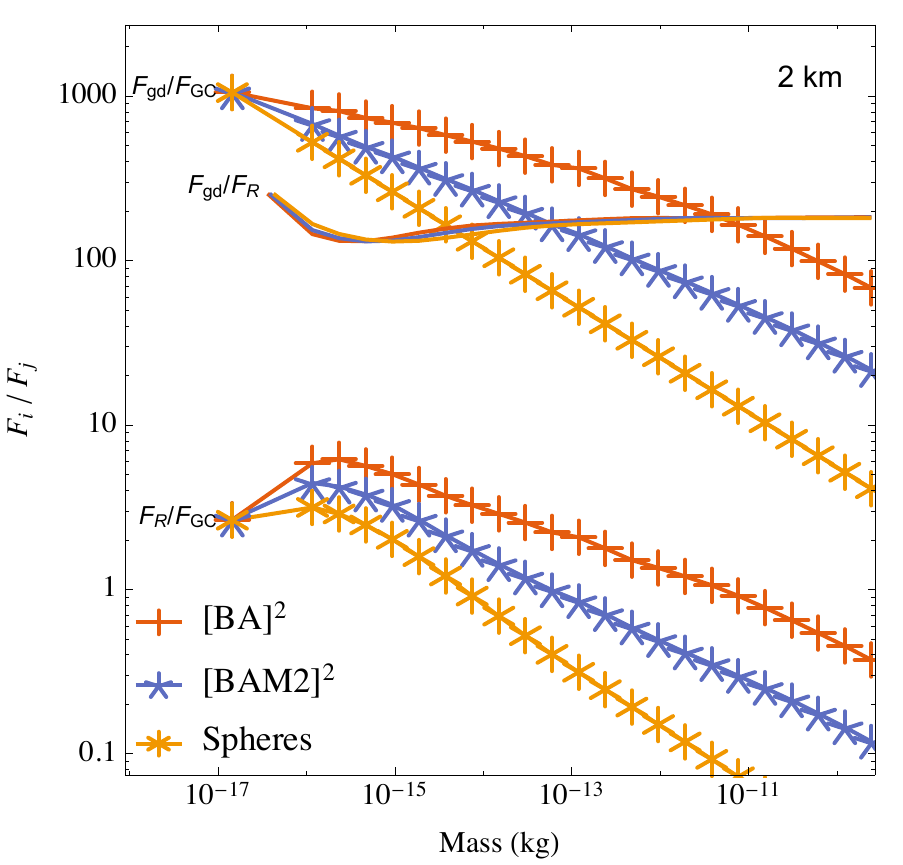}
\caption{Ratio of forces: $F_{gd}/F_{GC}$, $F_{gd}/F_R$ and $F_R/F_{GC}$ as a function of aggregate mass (or number of monomers) at 2 km from the comet center. Results for [BA]$^2$ and [BAM2]$^2$ aggregates, and solid spheres are shown.}
\label{fig:7}
\end{figure*}

\begin{figure*}[ht]
\centering\includegraphics[width=0.9\linewidth]{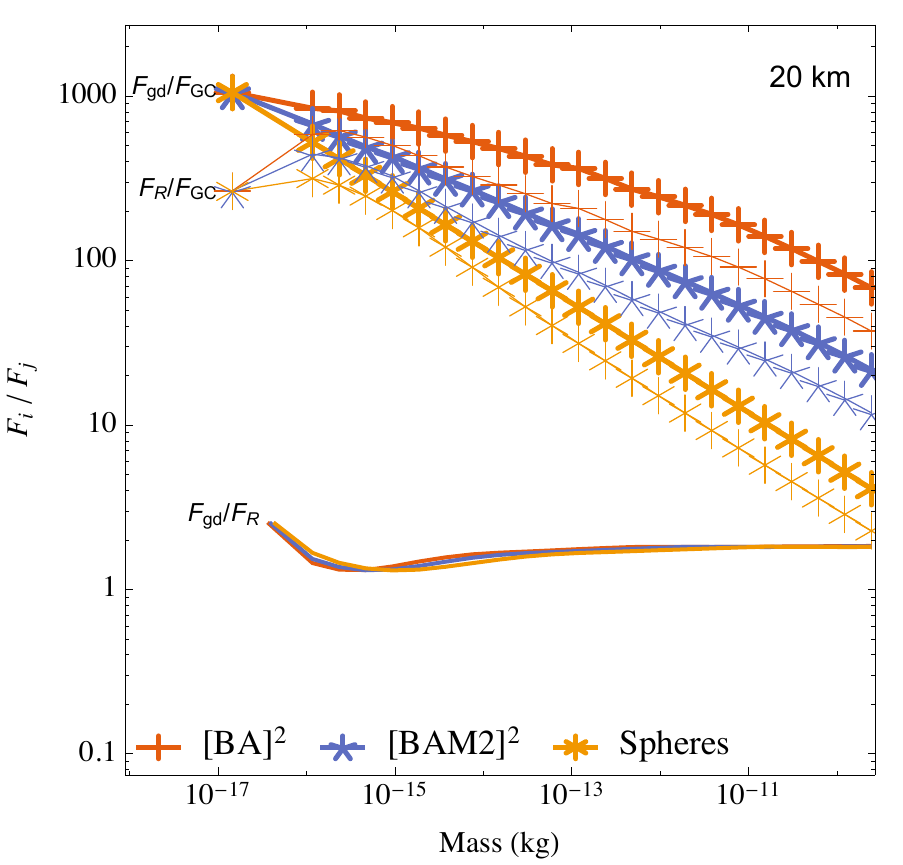}
\caption{Same as Fig.\ \ref{fig:7} but at a distance of 20 km from the comet center. $F_{gd}/F_{GC}$ and $F_R/F_{GC}$ were plotted using thicker and thinner lines to reduce confusion.}
\label{fig:8}
\end{figure*}

A quantitative comparison of the forces acting on a dust grain near the nucleus surface (Figs.\ \ref{fig:7} and \ref{fig:8}) clearly shows that gas drag dominates other effects by several orders of magnitude. The ratio of gas drag and radiation pressure, both proportional to the effective cross-section of dust aggregates, varies in the size range considered here, but is always much greater than unity. Gravitational forces, however, are proportional to the mass of the particle. For particles with fractal dimension $D_f>2$ ratio of particle mass to cross-section increases with size (for solid spheres it is linear with size). As a result, there is maximum size of ballistic particle that can be lifted by gas drag. The size threshold is a function of gas production, which is in turn a function of heliocentric distance. For our case (total gas production from a homogeneous spherical nucleus of about 2 kg s$^{-1}$) the heaviest particles that can be lifted dust are about $10^{-4}$ kg and $10^{-3}$ kg for BAM2 and BA structures respectively \citep{Gundlach2015}. Because the radiative force is much smaller than gas drag the force ratio becomes smaller than one for the particles with mass about $10^{-13}$-$10^{-12}$ kg (see Figs.\ \ref{fig:7} and \ref{fig:8}.) At the same time the radiative force exceeds the cometary gravity for the small particles which are the most effective light scatterers.  
 
At a distance of ten nucleus radii the situation is very different. Only the ratio of the gas drag to the force of cometary gravity remains the same, as in our model both the density of the gas (and as a result the gas drag) and the gravity of the nucleus vary inversely with the square of the distance to the nucleus. But at such large cometocentric distances gas drag is no longer dominant and becomes comparable to radiation pressure force (ratio $\sim 1$). The implication is that radiation pressure must be included when modeling dust dynamics beyond about 20 km above the surface. We note, however, that radiation pressure is strongly dependent on the optical characteristics of dust particles and more detailed treatment would require using effective media theory and/or more sophisticated computer methods.

\subsubsection{Dust velocity}

In Section \ref{InnerComaVelocity} we model the velocity of dust particles in the inner coma, which becomes of practical interest as Rosetta instruments probe for the first time the near-nucleus dynamics of dust particles. We focus on two observational phenomena: the velocity of dust particles registered by GIADA do not exceed $\sim$10 m s$^{-1}$ and depends only very weakly on particle mass \citep{DellaCorte}. Using these observational results, we try to provide useful information on the microscopic properties of the dust particles. 
 
As was shown above, near the nucleus dust is accelerated mainly by gas drag, proportionally to gas density and the square of the velocity difference between gas and dust (Eq. \ref{eq:CD}). In the planar surface approximation the gas density is constant and particles eventually attain terminal velocity. For a spherical nucleus, however, the $1/r^2$ drop in gas density implies a sharp decrease in gas drag with distance from the nucleus. Even the smallest dust particles attain only 5\% of the gas speed at about 10 nucleus radii. Note that we assume uniform sublimation across the spherical nucleus surface, implying a diluted gas surface density. If we assume that only a small fraction, $f_{active}$, of the nucleus surface is active, the surface gas density will increase in inverse proportion to $f_{active}$ leading to enhanced acceleration, but the dust will be accelerated only until the active region is displaced away by nucleus rotation.

As seen from the simulation results (see Fig.\ \ref{fig:9}), for the compact particles (solid spheres) even light particles ($~10^{-13}$ kg) retain the low speeds measured by GIADA. Dust particles 1000 times heavier are only half as fast. The range of velocities as a function of particle mass is smallest for solids grains and increases slightly with increasing of porosity. Absolute particle speeds also increase with increasing porosity: for the most massive particles, velocity is doubled with respect to solid spheres when the porosity of 70\% (BAM2) and is quadrupled with porosity of 85\% (BA). We note that all of these particles have a fractal dimension $D_f =3$. For particles with $D_f \sim2$ (BCCA) the dependence of dust speed on mass is insignificant. Comet dust is not expected to have $D_f \sim 2$, and our models corroborate this as they would be accelerated to speeds several times higher than what is observed. 
 
Our calculations have also shown that radiation pressure plays a minor role for near-nucleus dust dynamics. The difference between the case when the forces are added and the case when the forces are subtracted (sub-solar point) is less than 20\%.

\section{Conclusions}
\label{Conclusion}

We have presented a model of cometary dust capable of simulating the dynamics within the first few tens of km of the comet surface. Instruments aboard the Rosetta spacecraft show that the comet emits a range of dust particle sizes, from $\mu$m to mm, with porosities above 50\%. 

Here, we show that dust with these properties can be realistically simulated using hierarchical aggregates. We can simulate very large (mm) non-spherical particles and accurately determine their 1) effective cross-section and ratio of cross-section to mass, 2) gas drag coefficient, and 3) light scattering properties. Previously, such properties for the larger particles were impractical to calculate and required extrapolation from smaller sizes, which we show is inaccurate. We apply our model to the dynamics of dust in the vicinity of the nucleus of comet 67P and successfully reproduce the measured dust velocities {\color{red}} in the early stages of the Rosetta mission, when the comet was approximately 3.5 AU from the Sun. {\color {blue} However, the presented model does not provide a satisfactory quantitative explanation for the weak correlation seen between dust mass and dust velocity.}

At this stage, we employ a simple spherical comet nucleus, we model activity as constant velocity gas expansion from a uniformly active surface, and use Mie scattering. We discuss pathways to improve these simplifications in the future.

\section*{ Acknowledgements} 
This study was supported by DAAD (A/14/02432). Yu.S. and V.R.  wish  to  express  their  gratitude to the International Space Science Institute (ISSI) at Bern for support of their collaboration in the framework  of  the  Visiting  Science  Program.

\section*{References}
\label{References}

\clearpage


\begin{thebibliography}{00}

\bibitem[Armitage(2010)]{Armitage} Armitage P. J. 2010. Astrophysics of Planet Formation. Cambridge.

\bibitem[Baines et al.(1965)]{Baines} Baines, M. J., Williams, I. P., Asebiomo, A. S. 1965. Resistance to the motion of a small sphere moving through a gas. Mon. Not. R. Astr. Soc. 130, 63-74.
 
\bibitem[Bird(1994)]{Bird} Bird, G. A. 1994.  Molecular Gas Dynamics and the Direct Simulation of Gas Flows, 2nd edition. Oxford University Press, 484 pages. ISBN-13: 978-0198561958

\bibitem[Blum et al.(1996)]{Blum1996} Blum, J., Wurm, G., Kempf, 
S., Henning, T.\ 1996.\ The Brownian Motion of Dust Particles in the Solar 
Nebula: An Experimental Approach to the Problem of Pre-planetary Dust 
Aggregation.\ Icarus 124, 441-451.

\bibitem[Blum et al.(2014)]{Blum} Blum, J., Gundlach, B., M\"uhle, S., Trigo-Rodriguez, J. M. 2014. Comets formed in solar-nebula instabilities! - An experimental and modeling attempt to relate the activity of comets to their formation process. Icarus 235, 156-169.
 
\bibitem[Bohren \& Huffman(1983)]{Bohren} Bohren, C. F., Huffman, D. R., 1983. Absorption and Scattering of Light by Small Particles. Wiley, New York.
 
\bibitem[Bradley \& Brownlee(1986)]{Bradley} Bradley, J. P., Brownlee, D. E. 1986. Cometary particles: Thin sectioning and electron beam analysis. Science 231, 1542-1544.
 
\bibitem[Burns et al.(1979)]{Burns} Burns, J. A., Lamy, P. L., Soter, S., 1979. Radiation Forces on Small Particles in the Solar System. Icarus 40, 1-48.
 
\bibitem[Crifo(1987)]{Crifo} Crifo, J.F. 1987. Improved gas-kinetic treatment of cometary water sublimation and recondensation - Application to comet P/Halley. Astron.\ Astrophys.\ 187, 438-450.

\bibitem[Della Corte et 
al.(2015)]{DellaCorte} Della Corte, V., et al., 2015.\ GIADA: shining a light on the monitoring of the comet dust production from the nucleus of 67P/Churyumov-Gerasimenko.\ Astronomy and Astrophysics 583, A13. 

\bibitem[Draine \& Flatau(1994)]{Draine} Draine, B. T., Flatau, P. J. 1994. Discrete-dipole approximation for scattering calculations.  J. Opt. Soc. Am. A 11, 1491-1499.
 
\bibitem[Fulle et al.(2010)]{Fulle} Fulle, M., et al.,  2010. Comet 67P/Churyumov-Gerasimenko: the GIADA dust environment model of the Rosetta mission target.  Astron.\ Astrophys.\ 522, A63-A79.

\bibitem[Fulle et al.(2015)]{Fulle2015} Fulle, M., et al., 2015.\ Density and Charge of Pristine Fluffy Particles from 
Comet 67P/Churyumov-Gerasimenko.\ The Astrophysical Journal 802, L12. 

\bibitem[Fulle et al.(2015b)]{Fulle2015b}
Fulle, M., et al.,  2015.\ Rotating dust particles in the coma of comet 67P/Churyumov-Gerasimenko. Astronomy and Astrophysics 583, A14. 


\bibitem[Greenberg \& Hage(1990)]{Greenberg} Greenberg, J.M., Hage, J. I. 1990. From interstellar dust to comets - A unification of observational constraints. Astrophys.\ J.\ 361, 260-274.

\bibitem[Gulkis et al.(2015)]{Gulkis2015} 
Gulkis, S., et al.,   2015. Subsurface properties and early activity of comet 67P/Churyumov-Gerasimenko. Science, 10.1126 science.aaa0709.

\bibitem[Gundlach et al.(2015)]{Gundlach2015} Gundlach, B., Blum, J., Keller, H.~U., Skorov, Y.~V.\ 2015.\ What drives the dust activity of comet 67P/Churyumov-Gerasimenko?\ Astronomy and Astrophysics 583, A12.

\bibitem[Hadamcik et al.(2010)]{Hadamcik} Hadamcik, E.,  Sen, A. K., Levasseur-Regourd, A. C., Gupta, R., Lasue, J. 2010. Polarimetric observations of comet 67P/Churyumov-Gerasimenko during its 2008-2009 apparition. Astron.\ Astrophys.\ 517, A86-A94.

\bibitem[Ivanovski et al.(2015)]{Ivanovski2014}
Ivanovski, S., Zakharov, V., Della Corte, V., Lucarelli, F.,  Crifo, J. F.,  Rotundi, A., and M. Fulle. 2015. Aspherical rotating dust dynamics for GIADA experiment in the coma of 67P/Churyumov-Gerasimenko. Vol. 9, EPSC2014-629, 2014
European Planetary Science Congress.


\bibitem[Johansen et al.(2007)]{Johansen} Johansen, A., Oishi, J.S., MacLow, M.M., Klahr, H., Henning, T. 2007. Rapid planetesimal formation in turbulent circumstellar disks. Nature 448, 1022–1026.
 
\bibitem[Kelley et al.(2006)]{Kelley} Kelley, M.S., et al., 2006. A Spitzer Study of Comets 2P/Encke, 67P/Churyumov-Gerasimenko, and C/2001 HT50 (LINEAR-NEAT). Astrophys.\ J.\ 651, 1256-1271.
 
\bibitem[Kimura et al.(2006)]{Kimura} Kimura, H.,Kolokolova, L., Mann, I. 2006. Light scattering by cometary dust numerically simulated with aggregate particles consisting of identical spheres.  Astron.\ Astrophys.\ 449, 1243-1254.
 
\bibitem[Kissel et al.(1986)]{Kissel} Kissel, J. , et al., 1986. Composition of comet Halley dust particles from Giotto observations. Nature 321, 336-337.
 
\bibitem[K\"ohler et al.(2007)]{Kohler} K\"ohler, M., Minato, T., Kimura, H., Mann, I. 2007. Radiation pressure force acting on cometary aggregates. Adv. Space Res. 40, 266-271.
 
\bibitem[Kokhanovsky(2008)]{Kokhanovsky} Kokhanovsky, A. A. 2008. Aerosol Optics. Light Absorption and Scattering by Particles in the Atmosphere. Springer-Verlag Berlin Heidelberg.
 
\bibitem[Kolokolova et al.(2004)]{Kolokolova2004} Kolokolova, L., Hanner, M.~S., Levasseur-Regourd, A.-C., \& Gustafson, B.~{\AA}.~S.\ 2004.\  Physical properties of cometary dust from light scattering and thermal emission.\  Comets II 577 
 
\bibitem[Kolokolova et al.(2007)]{Kolokolova2007} Kolokolova, L., Kimura, H., Kiselev, N., Rosenbush, V. 2007. Polarimetric and infrared evidence of two types of dust in comets. Astron.\ Astrophys.\ 463, 1189-1196.

\bibitem[Kolokolova \& Kimura(2010)]{Kolokolova2010} Kolokolova, L., Kimura H. 2010. Comet dust as a mixture of aggregates and solid particles: model consistent with ground-based and space-mission results. Earth, Planets and Space 62, 17-22.

\bibitem[Kozasa et al.(1992)]{Kozasa} Kozasa, T., Blum, J., Mukai, T. 1992. Optical properties of dust aggregates. I - Wavelength dependence. Astron.\ Astrophys.\ 263, 423-432.

\bibitem[Laor \& Draine(1993)]{Laor} Laor, A., Draine, B. T. 1993. Spectroscopic Constraints on the Properties of Dust in Active Galactic Nuclei. Astrophys.\ J.\ 402, 441-468.


\bibitem[Lien(1990)]{Lien} Lien, D. J. 1990. Dust in comets. I - Thermal properties of homogeneous and heterogeneous grains. Astrophys.\ J.\ 355, 680-692.

\bibitem[Meakin(1987)]{Meakin1987} Meakin, P. 1987. Scaling properties for the growth probability measure and harmonic measure of fractal structures. Phys. Rev. A 35, 2234-2245.

\bibitem[Meakin et al.(1989)]{Meakin1989} Meakin, P., Donn, B., Mulholland, G.W. 1989. Collisions between point masses and fractal aggregates. Langmuir, 5, 510–518.

\bibitem[Minato et al.(2006)]{Minato2006} Minato, T., K{\"o}hler, M., Kimura, H., Mann, I., \& Yamamoto, T.\ 2006. Momentum transfer to fluffy dust aggregates from stellar winds. Astron.\ Astrophys.\ 452, 701 

\bibitem[Mishchenko(1996)]{Mishchenko} Mishchenko, M.I., Travis, L.D., Mackowski, D.W. 1996. T-matrix computations of light scattering by nonspherical particles: a review. Journal of Quantitative Spectroscopy and Radiative Transfer 55, 535-575.

\bibitem[Mukai et al.(1992)]{Mukai1992} Mukai, T., Ishimoto, H., Kozasa, T., Blum, J., Greenberg, J. M. 1992. Radiation pressure forces of fluffy porous grains. Astron.\ Astrophys.\ 262, 315-320.

\bibitem[Mukai \& Okada(2005)]{Mukai2005} Mukai, T., Okada, Y. 2005. Optical Properties of Large Aggregates. Workshop on Dust in Planetary Systems. Kauai, Hawaii. Editors: Krueger, H. and Graps, A., p.157-160.

\bibitem[Nakamura et al.(1994)]{Nakamura} Nakamura, R., Kitada, Y., Mukai, T. 1994. Gas drag forces on fractal aggregates. Planet.\ Space Sci.\ 42, 721-726.

\bibitem[Okuzumi et al.(2009)]{Okuzumi} Okuzumi, S., Tanaka, H., Sakagami, M.-a. 2009. Numerical Modeling of the Coagulation and Porosity Evolution of Dust Aggregates. Astrophys.\ J.\ 707, 1247-1263.

\bibitem[Petrova et al.(2000)]{Petrova} Petrova, E. V., Jockers, K., Kiselev, N. N. 2000. Light Scattering by Aggregates with Sizes Comparable to the Wavelength: An Application to Cometary Dust. Icarus 148, 2, 526-536.

\bibitem[Rotundi et al.(2015)]{Rotundi} Rotundi, A., et al., 2015. Dust measurements in the coma of comet 67P/Churyumov-Gerasimenko inbound to the Sun. Science 347, 6220, id. aaa3905.

\bibitem[Schulz et al.(2015)]{Schulz} Schulz, R., et al., 2015. Comet 67P/Churyumov-Gerasimenko sheds dust coat accumulated over the past four years. Nature 518, 216-218.

\bibitem[Shen et al.(2008)]{Shen} Shen, Y., Draine, B.T., Johnson, E.T.,  2008. Modeling Porous Dust Grains with Ballistic Aggregates. I. Geometry and Optical Properties. Astroph. J. 689, 260-275.

\bibitem[Shkuratov \& Grynko(2005)]{Shkuratov} Shkuratov, Y.G., Grynko, Y.S.,  2005. Light scattering by media composed of semitransparent particles of different shapes in ray optics approximation: consequences for spectroscopy, photometry, and polarimetry of planetary regoliths. Icarus 173, 16-28.

\bibitem[Shulman(1972)]{Shulman} Shulman, L.M., 1972. Dynamics of Cometary Atmospheres. Neutral Gas. Naukova Dumka, Kiev.

\bibitem[Sierks et al.(2015)]{Sierks} Sierks, H., et al., 2015. On the nucleus structure and activity of comet 67P/Churyumov-Gerasimenko. Science, Volume 347, Issue 6220, article id. aaa1044.

\bibitem[Skorov \& Rickman(1999)]{Skorov1999} Skorov, Yu.V., Rickman, H., 1999. Gas flow and dust acceleration in a cometary Knudsen layer. Planet.\ Space Sci.\ 47, 935-949.

\bibitem[Skorov et al.(2008)]{Skorov2008} Skorov, Yu.V., Keller, H.U., Rodin, A.V., 2008. Optical properties of aerosols in Titan's atmosphere. Planet.\ Space Sci.\ 56, 660-668.

\bibitem[Skorov et al.(2010)]{Skorov2010} Skorov, Yu.V., Keller, H.U., Rodin, A.V., 2010. Optical properties of aerosols in Titan's atmosphere: Large fluffy aggregates. Planet.\ Space Sci.\ 58, 1802-1810.

\bibitem[Skorov \& Blum(2012)]{Skorov2012} Skorov, Yu., Blum, J., 2012. Dust release and tensile strength of the non-volatile layer of cometary nuclei. Icarus 221, 1-11. 


\bibitem[Thomas et al.(2015)]{Thomas} Thomas, N., et al., 2015.  Evidence and Modelling of Dust Transport on the Nucleus of Comet 67P/Churyumov-Gerasimenko. 46th Lunar and Planetary Science Conference, held March 16-20, 2015 in The Woodlands, Texas. LPI Contribution No. 1832, 1712-1713.

\bibitem[van de Hulst(1957)]{vdHulst} van de Hulst, H.C., 1957. Light scattering by small particles. John Wiley \& Sons, New York.

\bibitem[Whipple(1950)]{Whipple} Whipple, F.L.A, 1950. Comet model. I. The acceleration of Comet Encke. Astrophys.\ J.\ 111, 375-394
 


 \end{thebibliography}
\end{document}